# The Ultimate Weapon for Ultra-Broadband 6G: Digital Beamforming and Doubly Massive mmWave MIMO

Bengt Lindoff, Carmen D'Andrea, Stefano Buzzi, Markus Törmänen and Per-Olof Brandt[1]

## ABSTRACT

The use of millimeter waves for wireless communications is one of the main technological innovations of 5G systems with respect to previous generations of cellular systems. Their consideration, however, has been up to now mainly restricted to the case in which analog or, at most, hybrid analog-digital beamforming structures were used, thus posing a limitation on the multiplexing capabilities and peak data rates that could be theoretically achieved at these frequencies. Recent progress in the field of electronics, however, has made the energy consumption of digital beamforming structures at least on par with that of analog beamforming, thus redeeming them from the ghetto they had been placed in over the last year. Digital beamforming, coupled with the use of large antenna arrays at both sides of the communication link, promises thus to be one of the secret weapons of future 6G networks, capable of unleashing unprecedented values of spectral and energy efficiency for ultra-broadband connectivity.

## INTRODUCTION

The adoption of millimeter wave (mmWave) carrier frequencies has been one of the main innovations introduced with the fifth generation of cellular systems. Traditionally, such frequencies have been considered inadequate to be used for wireless access in cellular systems, due to the large link lengths of early cellular generations and to the increased values of path loss at higher frequencies. Wireless researchers, however, attracted by the huge availability of unused spectrum in the above-6 GHz frequency range, re-discovered them about a decade ago [1]. It was indeed recognized that the path-loss was not so dramatic for link lengths of up to one-two hundred meters, and that it could be compensated by the use of antenna arrays with a sufficiently large number of active elements. MmWaves, even though having completely different characteristics from traditional sub-6 GHz cellular frequencies, could thus play a significant complementary role in cellular systems design [2]. In recent times, mmWaves for wireless access has thus become a very popular research area that has attracted interest from several researchers, both from academia and industry, and has been the focus of several research projects worldwide -- see, for instance [3]. One of the main research topics in the area has been that of beamformer design. Indeed, motivated by the fact that RF electronic circuitry at mmWave frequencies was not as efficient as it was at sub-6 GHz frequencies, and due to the impossibility of building in a cost-effective way transceivers with as many RF chains as the number of antennas, hybrid structures, with a number of RF chains significantly lower than the number of available antennas were studied [4, 5]. This has led to the design and analysis of analog beamformers (ABFs) and of hybrid analog digital beamformers (HBFs). Of course, such structures, while keeping low the system complexity, weaken the full potentialities that the use of large bandwidths at high frequencies could guarantee, by limiting the possibility of multiplexing a large number of users in the same time-frequency slot and by impeding the use of advanced precoding and postcoding schemes to suppress interference. This trade-off was highlighted some years ago for instance in [6, 7], where it was shown that digital beamforming schemes, although being more complex than analog and hybrid schemes, achieved better performance not only in terms of network throughput, but also in terms of energy efficiency, that is defined as the ratio of the overall network throughput over the consumed power, and is measured in bit/Joule. Nonetheless, digital beamforming schemes for mmWaves continued being ignored by mainstream research. Lately, the situation has changed, and digital beamforming at high frequencies has started attracting the attention it deserves [8, 9, 10].

---

[1] B. Lindoff and P.-O. Brandt are with BeammWave AB, Sweden. C. D'Andrea and S. Buzzi (corresponding author) are with the University of Cassino and Southern Latium, Italy, and with Consorzio Nazionale Interuniversitario per le Telecomunicazioni, Italy. M. Törmänen is with Lund University, Sweden.

We believe that digital beamforming, along with the use of large size multiple antenna arrays at both sides of the communication link (aka doubly massive MIMO), will be one of the key enablers of the ultra-large data rates predicted for future 6G networks.

This paper provides an updated overview of current technology for the realization of analog and fully digital beamformers, astonishingly showing that when integrating antennas, front-end filters and radio transceivers in a single RF chip, fully digital beamforming may consume even less power than simple analog beamformers. The paper also contains a numerical analysis of two relevant use cases for doubly massive MIMO system at millimeter waves, i.e., fixed wireless access (FWA) [11] and Vehicular-to-Infrastructure (V2I) communications [12], showing the superiority of digital beamforming, both in terms of system throughput and energy efficiency.

# RADIO ARCHITECTURES FOR MMWAVE BEAMFORMING

Beamforming of radio signals has found numerous applications in radar, radio astronomy and wireless communications. Radio implementations of beamforming can be based on a fully analog, fully digital beamforming or a combination, called hybrid beamforming. While hybrid beamforming may be a popular choice of architecture today, we focus the discussion below on analog and digital beamforming, in order to point out the fundamental differences in the design and its respective architecture's pros and cons.

As already discussed, traditional mmWave implementations use analog beamforming. In this case, beamforming is performed at the radio frequency (RF) through a bank of phase shifters, one per antenna element, and an analog power combiner (receiver) and power splitter (transmitter). Fig. 1 shows a principal sketch of a direct conversion analog beamforming architecture. This architecture only requires a single Local oscillator (LO) and mixer (at the transmitter and receiver), analog Low Pass Filter (LPF, transmitter and receiver), Voltage controlled Gain Amplifier (VGA, on the receiver side) and analog-to-digital converters (ADC) and digital-to-analog converters (DAC) at the receiver and transmitter, respectively, reducing the complexity. The antenna elements are typically clustered and implemented in an antenna panel.

While analog beamforming structures are believed to be power efficient, they are only capable of transmitting in one direction at a given time, which limits the potential of beamforming. For instance, analog beamforming cannot optimally cope with non-Line-of-Sight (nLoS) scenarios, i.e. scenarios where there is no direct radio part between the transmitter and receiver, and hence if there is a nLoS channel with multiple radio paths the beamformer need to choose a single direction to make the transmission in. Furthermore, the tracking abilities of transmission direction is degraded in case the mobile terminal (MT) is moving fast relative to the radio base station (BS) when using analog beamforming, since the beamforming direction needs to be decided first and then the radio measurement result of the analog signal combining can be determined in the digital decoder. This deficiency may cause quality of service degradation for services requiring ultra-low latency and very high data rates.

Furthermore, due to uncertainties in the phase shifters as well as limited dynamic and resolution in the phase shifter settings, there is also a performance loss in stationary cases compared to the beamforming using ideal phase shifters.

In contrast, a digital beamforming architecture performs the beamforming in the digital baseband processor. Fig. 2 shows a principal sketch of a direct conversion digital beamforming architecture. Each antenna has its own transceiver chain comprising a Local oscillator (LO) and mixer (transmitter and receiver), analog Low Pass Filter (LPF, transmitter and receiver), Voltage controlled Gain Amplifier (VGA, on the receiver side) and analog-to-digital converters (ADC) and digital-to-analog converters (DAC) at the receiver and transmitter, respectively. Performing the beamforming, or pre-coding in the digital domain enables the digital beamforming transceiver to simultaneously direct beams in theoretically infinite number of directions at a given time. Therefore, digital beamforming architectures can cope with nLoS scenarios and are also much better at tracking radio signals when mobile devices move fast relative to the radio BS, since information of all received signal streams is available at the digital decoder, and hence optimal beamforming can be made after the radio channel measurement. Furthermore, since the beamforming is made in the digital domain, optimal beamforming (pre-coders) can be achieved without implementation losses, compared to analog beamforming, and hence also in the stationary case, digital beamforming is in favor over analog beamforming.

# DIGITAL VS ANALOG BEAMFORMING – A POWER CONSUMPTION ANALYSIS

Analog beamforming has been believed to be a more power efficient solution than digital beamforming for mmWave transceivers, and the main argument has been that the digital solution with N antennas requires N ADC/DACs compared to the analog solution that only requires a single ADC/DAC. ADC/DAC has historically been consuming a lot of power especially if the number of bits is large and the sample rate is in the GHz range, needed when operating in mmWave band using bandwidths of several hundreds of MHz. Typically, ADC/DAC power consumption scales linearly with the sample rate, and exponentially with the number of bits used [13]. Therefore, there has been a consensus in academia and industry that a digital beamforming solution can barely compete with an analog one when operating in the mmWave frequency band when power consumption is the figure of merit to be considered.

However, this may have been the truth of yesterday, but – as will be further discussed in this section - with current state-of-the-art RF and AD/DA converter technologies, a digital beamforming solution as described below, can potentially achieve a power consumption on par or better than current analog versions.

In [9], a detailed power consumption analysis of digital and analog beamforming radio architectures for a mobile device use case was made. The analysis, that followed the approach used in [10], made a comparison of an analog beamforming transceiver assuming 8 antennas using current 5G-NR mobile device analog intermediate frequency radio solution towards an 8-antenna direct conversion digital beamforming solution, wherein the antennas, front-end filters and radio transceiver were integrated in a single RF chip.

The analysis concluded that with current RF and ADC/DAC state-of-the-art components, the power consumption for a digital beamforming architecture is at least on par with analog beamforming implementations.

The reason for this, maybe at a first glance quite surprising result, is twofold. First, the phase shifters and power combiners/splitters in the front-end radio, that is needed in the analog beamforming solution, has an insertion loss of around 8 dB, significantly increasing the front-end power consumption for analog beamforming. In particular, at the receiver side the LNAs need to compensate for the insertion loss in the phase shifters by increasing their gain, which means significantly higher LNA power consumption compared to the case of a digital beamforming solution. Accordingly, since the LNA power consumption scales with its gain, the LNA for an analog beamforming architecture consumes more than 6 times more power than the LNA required for a digital beamformer design in order to compensate for the 8 dB insertion loss in the phase shifters.

Second, the digital beamforming, with $N_{antenna}$ antennas, requires approximately $1.7\log_{10}(N_{antenna})$ fewer bits in the converters in order to achieve the same Signal to Quantization Distortion ratio (SQDR) as an analog beamforming solution, thus simplifying the AD/DA design for digital beamforming compared to the design required for analog beamforming.

The reason for this is that in a digital beamforming architecture, the ADC quantization noise is added on each respective receiver chain, and then in the digitally combiner, that coherently combines the signal from respective antenna, while the quantization noise, that is uncorrelated if the antenna spacing is at least half a wavelength, is non-coherently combined, and therefore quantization noise suppression of a factor $N_{antenna}$ is achieved. In the analog beamforming architecture, the signal is combined before the ADC and hence there is no such quantization noise interference suppression. Similar arguments hold for the transmitter chain and DAC case as well.

Accordingly, for an example system with 12 antennas, one 8-bit ADC/DAC pair is required in the analog case and twelve 6-bit ADC/DAC pairs are required in the fully digital case. Considering that the ADC/DAC power consumption grows exponentially with the number of bits, it is easily seen that the digital solution with 12 RF chains consumes only three times more power than the analog solution that has only one RF chain.

The above fact about the tougher radio front end design in the analog beamforming case together with the AD/DA design advantages for the digital solution, in combination with that the AD/DA converter technology development has come so far also at Gbits/s sampling rates. The ADC/DAC (at least if the number of bits used is 10 or less), indeed, does not consume significantly more power that other radio components today as can be seen from the power consumption numbers shown in Figs. 1 and 2. Hence, as shown in [9], the power consumption for an 8 antenna digital beamforming solution, in a typical operation mode, is around 580 mW for the transmitter and 250 mW for the receiver compared to 530 mW (transmitter) and 340 mW (receiver) for a current mobile device analog beamforming architecture.

## *POWER CONSUMPTION ESTIMATES FOR MMWAVE MASSIVE MIMO TRANSCEIVERS*

We now extend the results in [9] and analyze the power consumption for direct conversion analog, hybrid and digital beamforming architectures in a Massive MIMO context. The power consumption numbers for respective RF components analog and digital beamforming architectures are disclosed in Figs. 1 and 2. A hybrid beamforming architecture with $N_{RF}$, $1 < N_{RF} < N_{antennas}$, RF chains, still needs analog phase shifters, since the number of RF chains is fewer than the number of antennas. Therefore, the LNA power consumption will be as high as for the analog solution. However, due to multiple ADC/DACs some processing gain of quantization noise is achieved, and hence the number of ADC/DAC bits needed (and hence converter power consumption) will be in between the analog and digital beamforming architecture.

The communication is made on 28 GHz, assuming a system bandwidth of 200 MHz. For the analog beamforming case, the number of AD/DA bits is assumed to be 12/10 and 8/6 for the BS and mobile terminal respectively. For the hybrid beamforming the number of RF chains is assumed to be 3. For the digital and hybrid architecture, the number of AD/DA bits are reduced with $1.7\log_{10}(N_{antenna})$ and $1.7\log_{10}(N_{RF})$ respectively compared to the analog architecture, so the same converter SQDR for respective architectures is assumed. The total output power over all $N_{BS}$ antennas for the BS is assumed to be 30 dBm, while the total output power over all $N_{MT}$ antennas in a mobile terminal is assumed to be 18 dBm, implying the per PA output power will be in the range of 1-30 mW, depending on BS / mobile terminal implementation and number of antennas. In the figures, OFDM modulation is assumed in the communication, and due to its high peak-to-average power ratio a conservative PA efficiency of 15% is assumed.

Fig. 3 shows power consumption numbers for a BS and mobile terminal versus number of antennas, using current state-of-the-art RF and AD/DA converter technologies. As can be seen, the receiver power depends more on the number of antennas for the analog and hybrid solution compared to the digital due to the need of higher gain in LNA for compensating for phase shift insertion loss. Since the total transmitted output power is independent of the number of antennas (and hence, the output power per antenna is a decaying function of $N_{antenna}$), the receiver will then for some number of antennas consume more power than the transmitter. For mobile terminal/BS digital beamforming solution this happens at around 50/550 antennas, and for mobile terminal/BS analog and hybrid solution it happens at 12/170 antennas.

Focusing on the total power consumption, one can note that for a mobile terminal implementation, the power consumption for the digital beamforming solution is lower than for a corresponding analog and hybrid beamforming implementation already using 8-10 antennas.

For the BS implementation, the total transceiver power consumption is basically on par for analog and hybrid and slightly lower than for the digital beamforming architecture, regardless of the number of antennas used. However, as the number of antennas increases, the power consumption in the digital case becomes closer to that of the analog/hybrid solution, due to the need, in the analog and hybrid case, of more bits in ADC/DAC for achieving same SQDR, and due to the increased LNA power consumption to compensate the discussed insertion losses.

## USE CASES

It is yet not realistic, given current technology, assume that arrays with a large number of antennas coupled with a fully digital beamformer can be placed on a small-sized handheld mobile device. Maybe this will become possible in some years from now. However, there are already other relevant scenarios that can take advantage of such architectures.

One of these is certainly FWA. In areas where fiber-to-the-home is not available, the wide bandwidths available at mmWave make FWA a costly-effective solution with a performance comparable if not better than that of digital subscriber line (DSL) connections. In this case (see Fig. 4a), user devices may be placed outdoor on rooftops or on balconies, and it is reasonable to assume that there is a line-of-sight link with the BS antenna, an ideal setting for a doubly massive MIMO system operating at mmWave with digital beamforming.

Another key use case is that of V2I communications (see Fig. 4b). In this case the mobile transceiver is mounted on a car, so there is no concern about battery and/or antenna array space limitations. Placing BSs antennas on towers, and antennas on the roof of the car, it is possible to assume that there will be a LoS link with high likelihood. With multiple BSs coordinated multipoint schemes could also be implemented, so as to provide protection against possible signal blockages.

Wireless backhauling for distributed antenna systems [14] is another relevant use case. Mobile antennas can be embedded in access points, and mesh networking topologies could be implemented to optimize the connection towards the donor BS. This use case also refers to integrated terrestrial and aerial networks, where coordinated multipoint schemes aided by wireless backhaul can be envisaged with the aid of very low-earth-orbit nanosatellites and aerial platforms such as UAVs and Balloons [15].

*TWO SIMPLE NUMERICAL STUDIES*

To give a glimpse of the advantages granted by the joint use of doubly massive MIMO systems and fully digital beamforming at mmWave, we illustrate now sample results with reference to the FWA and V2I cases, assuming a 200 MHz communication bandwidth at 28 GHz carrier frequency. The power spectral density (PSD) of the noise is -174 dBm/Hz and the noise figure at the receiver is 5 dB. For the FWA scenario, the BS is positioned at a height of 30 m and the users are uniformly distributed in front of the BS in an angular region of 120° at distances between 10 and 300 m and heights uniformly distributed between 10 and 20 m. For the V2I scenario, the BS is positioned in the proximity of a crossroad at height of 15 m and the users are distributed at a fixed height of 1.65 m. In both the scenarios, the BS and users are equipped with planar arrays with half-wavelength antenna spacing. Figs. 5 and 6 report the downlink energy-efficiency versus throughput frontier for FWA and V2I scenarios, respectively. Energy consumption values reported in Figs. 1 and 2 and in [9] have been used to compute the system energy efficiency. Purely analog, beam-steering beamforming is compared with hybrid analog digital beamforming and with fully digital beamforming. For HBF and DBF, we consider both channel matched (CM) beamforming and zero-forcing (ZF) beamforming. Results clearly show the superiority of DBF, which is capable of granting, over a 200 MHz bandwidth, aggregated throughput of tens of Gbps, and, at the same time, much larger values of energy-efficiency.

# CONCLUSIONS: TRENDS FOR MASSIVE MIMO MMWAVE RADIO ARCHITECTURES TOWARDS 2030

From the above discussions we have seen that the semiconductor technology and chip integration for mmWave transceivers has come so far that digital beamforming already today is on par or better than analog and hybrid beamforming when it comes to power consumption as long as one can integrate the antennas, front-end filters and radio transceiver in a single RF chip. The integration is possible as long as the output transmission power stays below around 15 dBm per transmitter, which also have been shown above, is the case in mmWave Massive-MIMO applications. We can also see from the per RF component power consumption that the ADC/DAC, as long as the number of bits is less than 10, does not significantly consume more power than other RF components. In further power consumption reduction research for mmW transceivers, it might be important to also investigate methods for power reductions in the LO design, since phase noise suppression seems to be challenging at mmWave frequency, making the LO power consumption, in fact, larger than for corresponding 8-bit ADC/DAC.

A numerical study based on two relevant use cases, namely FWA and V2I communications, has shown that digital beamforming largely outperforms analog and hybrid beamforming schemes, achieving a much better trade-off of system throughput versus energy-efficiency.

We think that this implies that analog as well as hybrid beamforming solutions, that have been seen as the current optimal trade-off transceiver architecture for mmWave Massive MIMO, will in the next few years be phased out in favor of fully digital beamforming solutions, with an ever-increasing size of the antenna arrays, not only at the BS, but also at the user devices.

# ACKNOWLEDGEMENTS
The work of C. D'Andrea and S. Buzzi has been supported in part by the MIUR Program "Dipartimenti di Eccellenza 2018-2020" and in part by the MIUR PRIN 2017 Project "LiquidEdge".

# REFERENCES

[1] T. S. Rappaport *et al*., "Millimeter Wave Mobile Communications for 5G Cellular: It Will Work!," in *IEEE Access*, vol. 1, pp. 335-349, 2013, doi: 10.1109/ACCESS.2013.2260813.

[2] E. Bjornson, L. Van der Perre, S. Buzzi and E. G. Larsson, "Massive MIMO in Sub-6 GHz and mmWave: Physical, Practical, and Use-Case Differences," in *IEEE Wireless Communications*, vol. 26, no. 2, pp. 100-108, April 2019, doi: 10.1109/MWC.2018.1800140.

[3] M. Tercero *et al*., "5G systems: The mmMAGIC project perspective on use cases and challenges between 6–100 GHz," *2016 IEEE Wireless Communications and Networking Conference Workshops (WCNCW)*, Doha, Qatar, 2016, pp. 200-205, doi: 10.1109/WCNCW.2016.7552699.

[4] S. Han, C. I, Z. Xu and C. Rowell, "Large-scale antenna systems with hybrid analog and digital beamforming for millimeter wave 5G," in *IEEE Communications Magazine*, vol. 53, no. 1, pp. 186-194, January 2015, doi: 10.1109/MCOM.2015.7010533.

[5] R. Méndez-Rial, C. Rusu, N. González-Prelcic, A. Alkhateeb and R. W. Heath, "Hybrid MIMO Architectures for Millimeter Wave Communications: Phase Shifters or Switches?," in IEEE Access, vol. 4, pp. 247-267, 2016, doi: 10.1109/ACCESS.2015.2514261.

[6] S. Buzzi and C. D'Andrea, "Are mmWave Low-Complexity Beamforming Structures Energy-Efficient? Analysis of the Downlink MU-MIMO," 2016 IEEE Globecom Workshops (GC Wkshps), Washington, DC, 2016, pp. 1-6, doi: 10.1109/GLOCOMW.2016.7848841.

[7] S. Buzzi and C. D'Andrea, "Energy efficiency and asymptotic performance evaluation of beamforming structures in doubly massive MIMO mmWave systems," in IEEE Transactions on Green Communications and Networking, vol. 2, no. 2, pp. 385-396, June 2018, doi: 10.1109/TGCN.2018.2800537.

[8] M. Kappes, "All-digital antennas for mmWave systems," Microwave Journal, pp. 8-11, September 2019.

[9] "Digital Beamforming for mobile devices – the power efficient architecture for 5G on mmWave," BeammWave White Paper, September 2020 (available at https://www.beammwave.com/whitepapers). Accessed on date: 26 Feb. 2021.

[10] S. Dutta, C. N. Barati, D. Ramirez, A. Dhananjay, J. F. Buckwalter and S. Rangan, "A Case for Digital Beamforming at mmWave," in IEEE Transactions on Wireless Communications, vol. 19, no. 2, pp. 756-770, Feb. 2020, doi: 10.1109/TWC.2019.2948329.

[11] C. Chen, O. Kedem, C. R. C. M. d. Silva and C. Cordeiro, "Millimeter-Wave Fixed Wireless Access Using IEEE 802.11ay," in *IEEE Communications Magazine*, vol. 57, no. 12, pp. 98-104, December 2019, doi: 10.1109/MCOM.001.1900076.

[12] D. Kong, J. Cao, A. Goulianos, F. Tila, A. Doufexi and A. Nix, "V2I mmWave Connectivity for Highway scenarios," *2018 IEEE 29th Annual International Symposium on Personal, Indoor and Mobile Radio Communications (PIMRC)*, Bologna, Italy, 2018, pp. 111-116, doi: 10.1109/PIMRC.2018.8580735.

[13] B. Murmann, "ADC Performance Survey 1997-2020 (ISSCC & VLSI Symposium)", [Online]. Available: http://web.stanford.edu/~murmann/adcsurvey.html. Accessed on date: 26 Feb. 2021.

[14] M. N. Islam, S. Subramanian and A. Sampath, "Integrated Access Backhaul in Millimeter Wave Networks," *2017 IEEE Wireless Communications and Networking Conference (WCNC)*, San Francisco, CA, 2017, pp. 1-6, doi: 10.1109/WCNC.2017.7925837.

[15] M. Giordani and M. Zorzi, "Non-Terrestrial Networks in the 6G Era: Challenges and Opportunities," in *IEEE Network*, doi: 10.1109/MNET.011.2000493.

# BIOGRAPHIES

**Bengt Lindoff**, IEEE Senior Member, received his MSc EE in 1992 and PhD in Mathematical Statistics in 1997 from Lund University, Sweden. Bengt has 25 years of experience of wireless communication research within Ericsson and Huawei. Currently he is Chief Systems Architect at BeammWave AB, Lund Sweden, a start-up designing disruptive RF chip solutions for mmWave 5G-NR. Bengt has authored/co-authored more than 30 journal and conference papers, and is inventor to over 500 filed patents and he is recognized as one of the world's most prolific inventors within the area of wireless communication.

**Carmen D'Andrea** (S'18, M'20) was born in Caserta, Italy on 16 July 1991. She received the B.S., M.S. and the Ph.D degrees, all with honors, in Telecommunications Engineering from the University of Cassino and Southern Latium in 2013, 2015, and 2019 respectively. She spent short visiting research periods with the Wireless Communications (WiCom) Research Group in the Department of Information and Communication Technologies at Universitat Pompeu Fabra in Barcelona, Spain, and with the Communication System Division of the Department of Electrical Engineering at the Linkoping University, Sweden. She is currently Assistant Professor at the University of Cassino and Southern Latium and her research interests are focused on wireless communication and signal processing, with current emphasis on mmWave communications and massive MIMO systems.

**Stefano Buzzi** (M'98-SM'07) is Full Professor at the University of Cassino and Lazio Meridionale, Italy. He received the M.Sc. degree (summa cum laude) in Electronic Engineering in 1994, and the Ph.D. degree in Electrical and Computer Engineering in 1999, both from the University of Naples "Federico II". He is a former Associate Editor of the IEEE Signal Processing Letters and of the IEEE Communications Letters, has been the guest editor of three IEEE JSAC special issues (June 2014, April 2016, and April 2019), and is a former Editor for the IEEE Transactions on Wireless Communications. Dr. Buzzi's research interests are in the broad field of communications and signal processing, with emphasis on wireless communications. He has co-authored about 160 technical peer-reviewed journal and conference papers, and among these, the highly-cited survey paper "What will 5G be?" (IEEE JSAC, June 2014) on 5G wireless networks.

**Markus Törmänen** (S'06–M'10–SM'12) received the PhD degree in Circuit Design in 2010 from Lund University, Sweden. He was an Assistant Professor at Lund University 2010-2013, and since 2014 he has been Associate Professor and Docent at Lund University. He has authored/co-authored more than 70 international peer reviewed journal and conference papers. He is the Program Director for Electrical Engineering (M.Sc. Eng) at Lund University and has been awarded the IEEE Senior Member grade. His research interests include design of analog/RF and mm-wave circuits. He is also part of the RF group designing the low-level RF system for the European Spallation Source (ESS) and co-founder of BeammWave AB.

**Per-Olof Brandt**, MSc EE (1993), Lund University, Sweden. Per-Olof has 27 years' experience of RF design at Ericsson, Cambridge Silicon Radio and Imagination/Kisel. Per-Olof is co-founder and Chief Technical Officer at BeammWave AB, Lund Sweden, design lead for RFIC and HW design. Per-Olof has over 20 patents and has held positions as Senior Specialist as well as Consultant in the fields of RFIC, Power Amplifier and antenna-interface design.

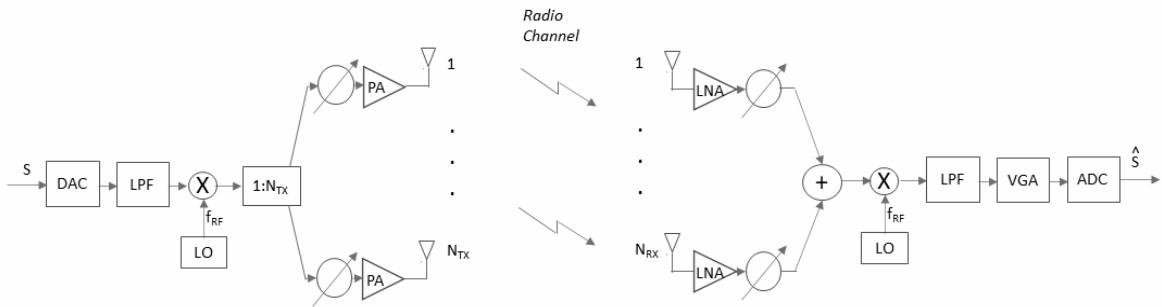

*Figure 1*. Analog beamforming radio architecture. After the DAC and low pass filter, the transmitted baseband signal S is converted to a radio signal in a single up-conversion mixer and thereafter split in a power splitter and fed to a bank of phase shifters that are used for performing beamforming. The beamformed signals are then amplified in the Power Amplifiers (PA), prior to be transmitted via the respective $N_{TX}$ transmit antennas (left). At the receiver side (right) the respective received signal is first amplified in a Low Noise Amplifier (LNA) and fed to a phase shifter and the beamformed received signals are combined and then down-converted in a single mixer to a baseband signal, that is low pass filtered and amplified prior to the ADC. This architecture uses just one pair of up and down converting and ADC and DAC at the baseband. Typical values for the power consumption for the different RF components for a state-of-the-art mmWave massive MIMO are 43/3 mW for the DAC (BS/MT), 30 mW for the LO both at the transmitter and receiver side, IL for the phase shifters is 8 dB, 1-30 mW for the PA with 15% efficiency, 0.5 mW and 1.6 mW for the LPF at the transmitter and receiver, respectively, 36 mW for the LNA, 1.3 mW for the VGA and 172/11 mW for the ADC (BS/MT). The ADC/DAC power consumption is different for BSs and MTs due to different bit resolution requirements.

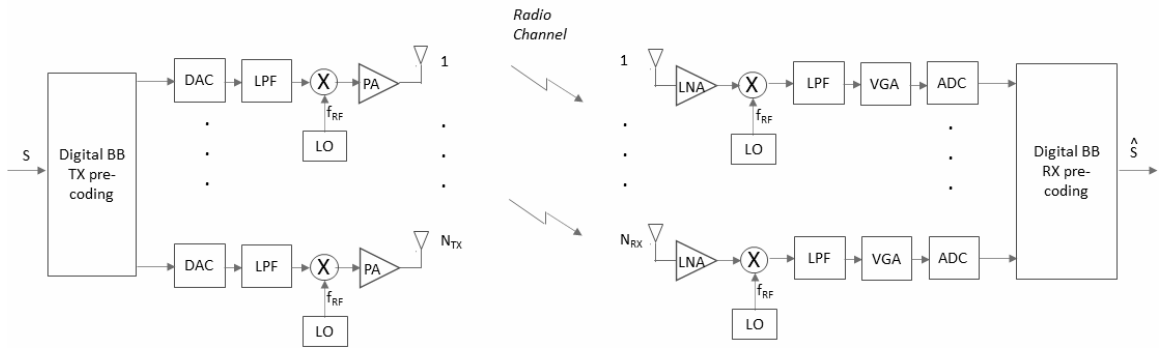

*Figure 2. Digital beamforming radio architecture, where each antenna element has a single transceiver chain. The transmitted signal S is beamformed (pre-coded) for each transmitter chain in the digital baseband prior to the DAC (left). At the receiver, the beamforming (pre-coding) is performed after the ADC in the digital baseband (right). Typical values for the power consumption for the different RF components for a state-of-the-art mmWave massive MIMO are 43/3 mW divided by a factor taking into account the number of antennas for the DAC (BS/MT), 30 mW for the LO both at the transmitter and receiver side, 1-30 mW for the PA with 15% efficiency, 0.5 mW and 1.6 mW for the LPF at the transmitter and receiver, respectively, 5.6 mW for the LNA, 1.3 mW for the VGA and 172/11 mW divided by a factor taking into account the number of antennas for the ADC (BS/MT). ADC/DAC power consumption is different for BSs and MTs due to different bit resolution requirements. In order to achieve the same number of Signal to Quantization Distortion ratio (SQDR) for the digital solution as the analog solution the number of bits can be reduced based on the number of antennas.*

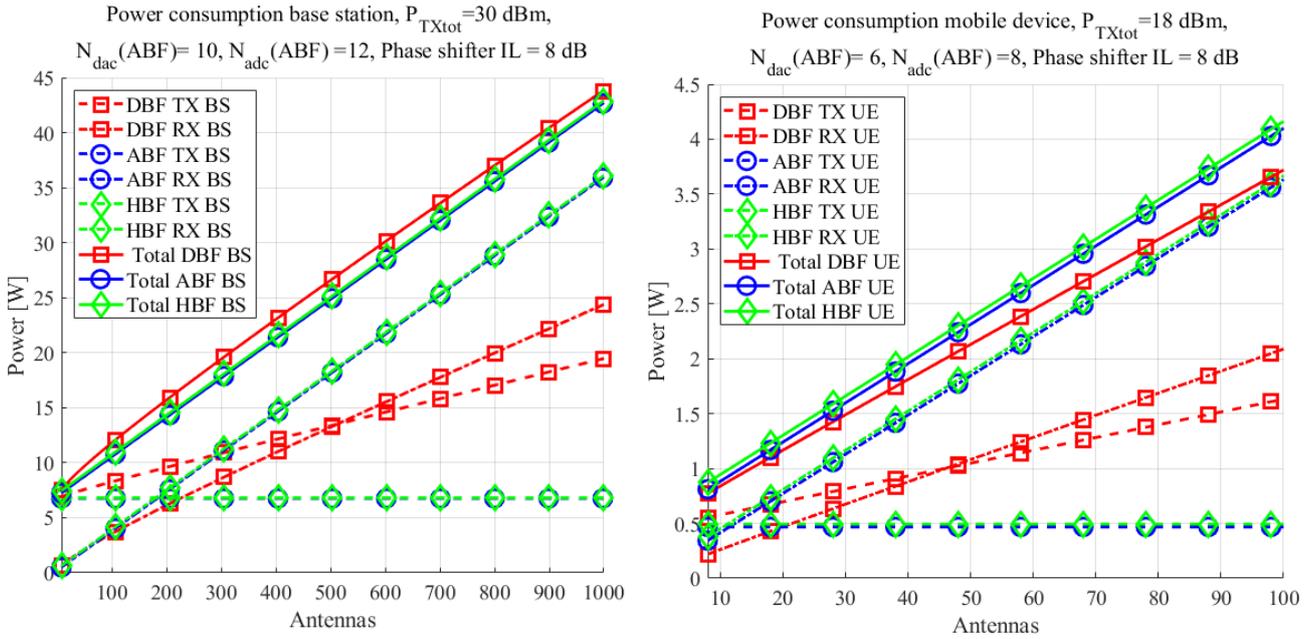

***Figure 3.*** *Power consumption estimates for mmWave massive-MIMO BS (left) and mobile terminal (right) using analog (ABF), hybrid (HBF) and digital (DBF) beamforming radio architectures assuming current state-of-the-art RF and AD/DA converter technologies [9]. As can be seen, the total transceiver power consumption is basically on par for the BS analog, hybrid and digital beamforming architecture, while it is lower for a digital beamforming compared to analog beamforming in a mobile terminal implementation.*

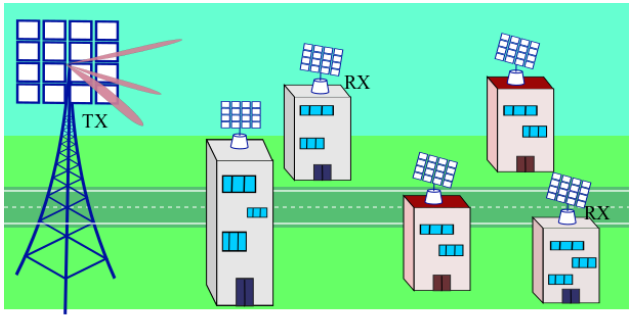
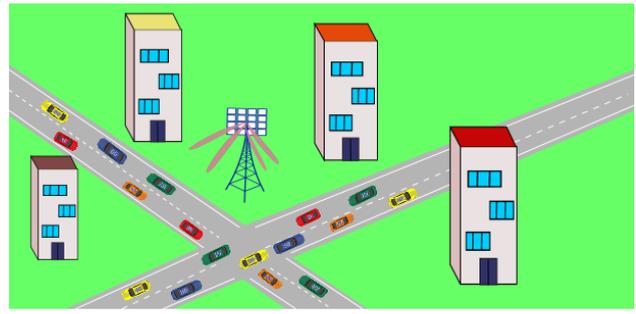

a) FWA scenario  b) V2I scenario

**Figure 4.** *Figurative representation of the considered FWA and V2I scenarios.*

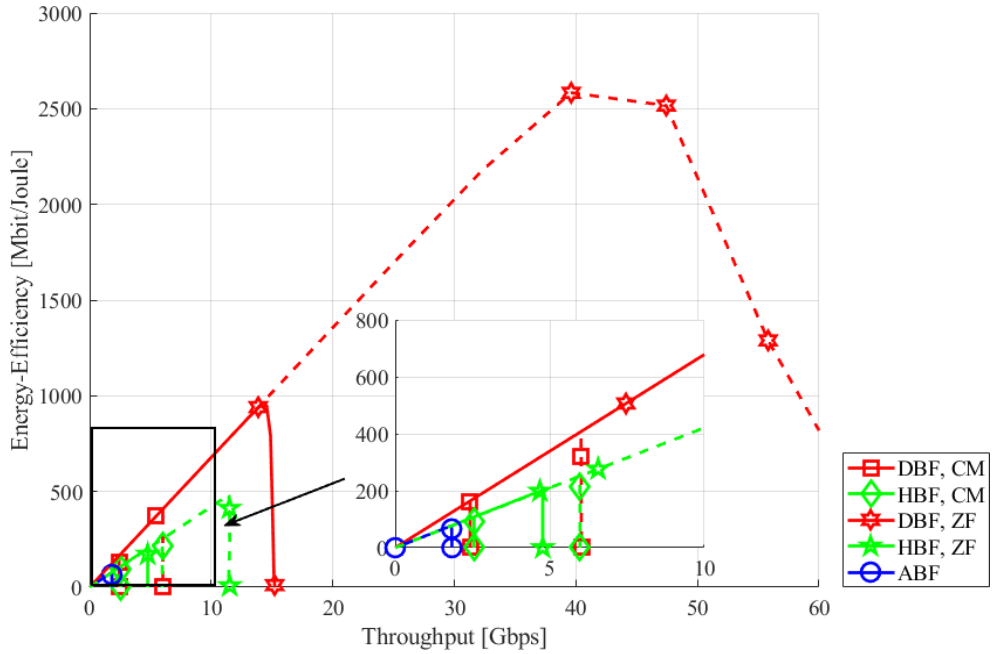

*Figure 5.* Energy-efficiency vs. Throughput on the downlink in the FWA scenario, with DBF, HBF and ABF. We assume a number of receivers simultaneously served by the BS equal to 10, and 64 antennas at both sides of the communication links. Solid lines refer to the case of one transmitted stream per TX-RX link, while dashed lines refer to the case of two streams. The number of RF chains in the HBF is assumed equal to the number of streams multiplied by the number of receivers at the BS and equal to the number of streams at the receivers.

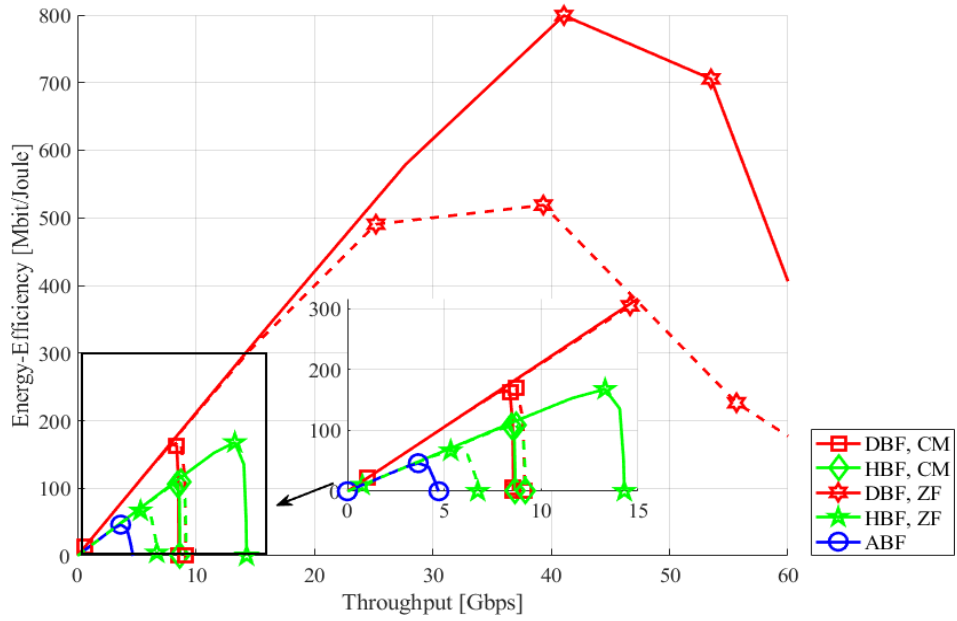

*Figure 6.* *Energy-efficiency vs. Throughput on the downlink in the V2I scenario, with DBF, HBF and ABF. We assume a number of cars simultaneously served by the BS equal to 32, 256 antennas at the BS, and 64 at the receivers. Solid lines refer to the case of one transmitted stream per TX-RX link, while dashed lines refer to the case of two streams. The number of RF chains in the HBF is assumed equal to the number of streams multiplied by the number of receivers at the BS and equal to the number of streams at the receivers.*